\documentclass[twocolumn,aps,prl]{revtex4-1}
\usepackage[usenames,dvipsnames]{xcolor}
\usepackage{amsmath}
\usepackage{amsfonts}
\usepackage{amssymb}
\usepackage[colorlinks=true,citecolor=blue,linkcolor=red]{hyperref}
\usepackage{graphicx}
\usepackage{bbold}					% for \mathbb{1} as a nice unit matrix symbol
\usepackage[makeroom]{cancel}		% crossed terms in tables
\usepackage{multirow}				% merge cells vertically in tables
\usepackage[normalem]{ulem}        % strike-through text
\usepackage{array}
\usepackage{pdfpages}
\makeatletter
\AtBeginDocument{\let\LS@rot\@undefined}
\makeatother

	\newcolumntype{x}[1]{>{\centering\let\newline\\\arraybackslash\hspace{0pt}}p{#1}}
	
	\renewcommand{\Re}{\operatorname{Re}}
	\renewcommand{\Im}{\operatorname{Im}}

	\DeclareMathAlphabet{\mathbbold}{U}{bbold}{m}{n}
	\newcounter{subeqn} %
	\makeatletter
	\@addtoreset{subeqn}{equation}
	\makeatother

\definecolor{TB}{rgb}{0,0,0} %uncomment for all black

\begin{document}

\title{Non-Hermitian bulk-boundary correspondence and auxiliary generalized Brillouin zone theory}

\author{Zhesen Yang$^{1,2}$}\thanks{These two authors contributed equally}
\author{Kai Zhang$^{1,2}$}\thanks{These two authors contributed equally}
\author{Chen Fang$^{1}$}\email[Corresponding author: ]{cfang@iphy.ac.cn}
\author{Jiangping Hu$^{1,2,3,4}$}\email[Corresponding author: ]{jphu@iphy.ac.cn}

\affiliation{$^{1}$Beijing National Laboratory for Condensed Matter Physics,
	and Institute of Physics, Chinese Academy of Sciences, Beijing 100190, China}
\affiliation{$^{2}$University of Chinese Academy of Sciences, Beijing 100049, China}
\affiliation{$^{3}$Kavli Institute for Theoretical Sciences and CAS Center for Excellence in Topological Quantum Computation, University of Chinese Academy of Sciences, Beijing 100190, China}
\affiliation{$^{4}$South Bay Interdisciplinary Science Center, Dongguan, Guangdong Province, China}
	
\begin{abstract}
We provide a systematic and self-consistent  method to calculate the generalized Brillouin Zone (GBZ) analytically in one dimensional non-Hermitian systems, which helps us to understand the non-Hermitian bulk-boundary correspondence. In general, a $n$-band non-Hermitian Hamiltonian is constituted by $n$ distinct sub-GBZs, each of which is a piecewise analytic closed loop. Based on the concept of resultant, we can show that all the analytic properties of the GBZ can be characterized by an algebraic equation, the solution of which in the complex plane is dubbed as auxiliary GBZ (aGBZ). We also provide a systematic method to obtain the GBZ from aGBZ. Two physical applications are also discussed. 
Our method provides an analytic approach to the spectral problem of open boundary non-Hermitian systems in the thermodynamic limit.
\end{abstract}
	
\maketitle

{\em Introduction.}---Bulk-boundary correspondence (BBC) has played a fundamental role in the development of topological band theory~\cite{Hasan2010,Qi2011,Bernevig2013}. For example, the chiral edge state can be faithfully predicted by the Chern number. A hidden assumption of the celebrated BBC is that the bulk properties of the open boundary condition (OBC) Hamiltonian can be well approximated by the Bloch Hamiltonian with periodic boundary condition (PBC)~\cite{Cobanera2016}. However, this hidden assumption is challenged in some non-Hermitian systems recently~\cite{yaoEdgeStatesTopological2018b,yaoNonHermitianChernBands2018b,PhysRevLett.123.170401,PhysRevLett.123.246801,kunstBiorthogonalBulkBoundaryCorrespondence2018a,yokomizoNonBlochBandTheory2019a,zhangCorrespondenceWindingNumbers2019,Okuma2020,xiongWhyDoesBulk2018c,martinezalvarezNonHermitianRobustEdge2018b,leeAnatomySkinModes2019b,Longhi2019, herviouDefiningBulkedgeCorrespondence2019,zirnsteinBulkboundaryCorrespondenceNonHermitian2019a,JiangHui2019,longhiProbingNonHermitianSkin2019,PhysRevB.100.165430,1907.11562,Xiao2019,1908.02759,ZhangXZ2019,1910.03229,Foa_Torres_2019,Borgnia2020,brzezickiHiddenChernNumber2019,dengNonBlochTopologicalInvariants2019a,edvardssonNonHermitianExtensionsHigherorder2019a,ezawaNonHermitianBoundaryInterface2019a,kunstNonHermitianSystemsTopology2019b,wangNonHermitianNodallineSemimetals2019b,KawabataPRX,Gong2018,LiLinhu2019,ShenHT2018,Okuma2019,JinL2019,Takata2018,Longhi2020,Ghatak2019c,MartinezAlvarez2018c,Zeuner2015,Rudner2009,Lieu2018,Brody2014c,2020arXiv200302219Y}. To be more precise, when the OBC Hamiltonian has non-Hermitian skin effect~\cite{yaoEdgeStatesTopological2018b,yaoNonHermitianChernBands2018b,PhysRevLett.123.170401,PhysRevLett.123.246801,kunstBiorthogonalBulkBoundaryCorrespondence2018a,yokomizoNonBlochBandTheory2019a,zhangCorrespondenceWindingNumbers2019,Okuma2020,xiongWhyDoesBulk2018c,martinezalvarezNonHermitianRobustEdge2018b,leeAnatomySkinModes2019b,Longhi2019, herviouDefiningBulkedgeCorrespondence2019,zirnsteinBulkboundaryCorrespondenceNonHermitian2019a,JiangHui2019,longhiProbingNonHermitianSkin2019,PhysRevB.100.165430,1907.11562,Xiao2019,1908.02759,ZhangXZ2019,1910.03229}, the spectrum between OBC and PBC can be totally distinct~\cite{yaoEdgeStatesTopological2018b,yaoNonHermitianChernBands2018b,PhysRevLett.123.170401,PhysRevLett.123.246801,kunstBiorthogonalBulkBoundaryCorrespondence2018a,yokomizoNonBlochBandTheory2019a,zhangCorrespondenceWindingNumbers2019,Okuma2020,xiongWhyDoesBulk2018c}. It has been revealed that much important information of the OBC Hamiltonian can be encoded from the generalized Brillouin zone (GBZ)~\cite{yaoEdgeStatesTopological2018b,yaoNonHermitianChernBands2018b,yokomizoNonBlochBandTheory2019a,zhangCorrespondenceWindingNumbers2019}, which is a generalization of Brillouin Zone (BZ) under the OBC in both {\em Hermitian} and {\em non-Hermitian} systems. Although the OBC breaks the translational symmetry and the generalization of BZ seems odd, the basic idea of GBZ is to find a suitable generalized Bloch Hamiltonian (GBZ Hamiltonian) such that the boundary scattering can be regarded as a perturbation. Thus the calculation of GBZ becomes important and has drawn extensive attentions recently~\cite{leeAnomalousEdgeState2016a,leykamEdgeModesDegeneracies2017e,xiongWhyDoesBulk2018c,martinezalvarezNonHermitianRobustEdge2018b,yaoEdgeStatesTopological2018b,yaoNonHermitianChernBands2018b,kunstBiorthogonalBulkBoundaryCorrespondence2018a,martinezalvarezTopologicalStatesNonHermitian2018b,leeAnatomySkinModes2019b,yokomizoNonBlochBandTheory2019a,herviouDefiningBulkedgeCorrespondence2019,zirnsteinBulkboundaryCorrespondenceNonHermitian2019a,JiangHui2019,PhysRevLett.123.170401,PhysRevLett.123.246801,longhiProbingNonHermitianSkin2019,Longhi2019,Okuma2020,zhangCorrespondenceWindingNumbers2019,Borgnia2020,brzezickiHiddenChernNumber2019,dengNonBlochTopologicalInvariants2019a,edvardssonNonHermitianExtensionsHigherorder2019a,ezawaNonHermitianBoundaryInterface2019a,ghatakNewTopologicalInvariants2019,kunstNonHermitianSystemsTopology2019b,wangNonHermitianNodallineSemimetals2019b}. Unfortunately, up to now, there is no universal analytical method to calculate the GBZ, and the numerical method is not only time-consuming but also unreliable due to the existence numerical errors that are extremely sensitive to the lattice size and calculation precision~\cite{SM2,colbrookHowComputeSpectra2019b,bottcher2005,REICHEL1992153}.
In this paper, we solve this challenging problem analytically based on the concept of {\em auxiliary GBZ} (aGBZ). We show that the GBZ of a $n$-band Hamiltonian has $n$ distinct sub-GBZs, corresponding to the $n$ distinct bands. Each sub-GBZ is a piecewise analytic closed loop, and can be described by a common algebraic equation, namely, the aGBZ equation, which can be calculated based on the concept of resultant of polynomials~\cite{Resultant2019,woodyDepartmentMathematicsComputer,jansonRESULTANTDISCRIMINANTPOLYNOMIALS,SM2}. We also provide a systematic method to pick up the GBZ from aGBZ. As applications of our method, we discuss the perturbation-failure effect and the BBC in the case where each band has its respective, distinct sub-GBZ.

\begin{figure}[t]
	\centerline{\includegraphics[height=2.2cm]{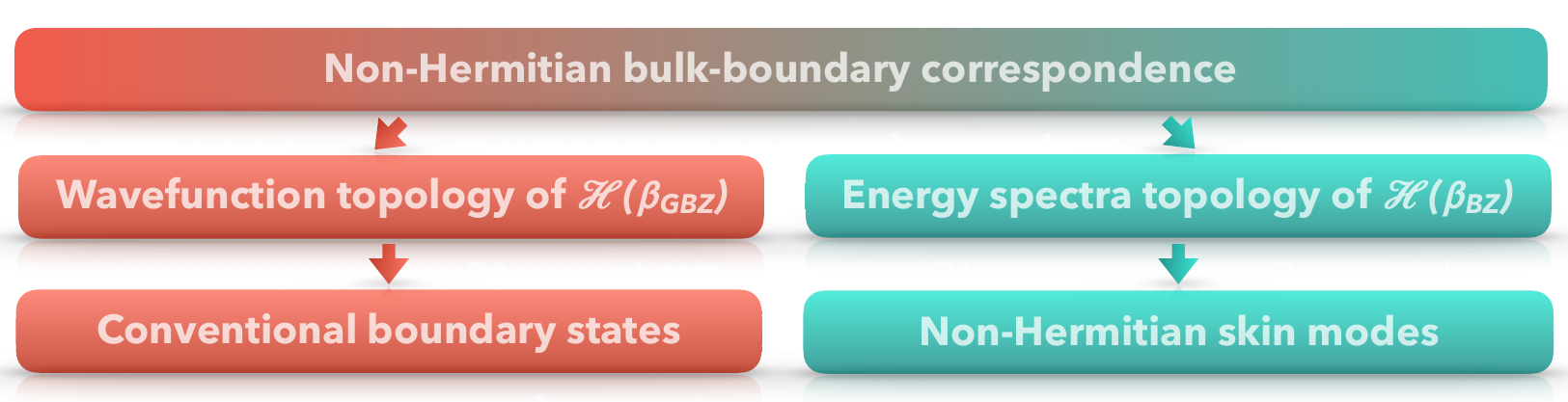}}
	\caption{The non-Hermitian bulk-boundary correspondence has two-fold meaning.
	\label{F1}}
\end{figure}

{\em BBC and GBZ.}---We start from the following one-dimensional (1D) Bloch Hamiltonian
\begin{equation}
\mathcal{H}(k)=\mathcal{H}_0(k)+i\lambda\mathcal{H}_{1}(k),\quad \lambda\in \mathbb{R},
\label{E0}
\end{equation}
where $\mathcal{H}_{0/1}(k)=\mathcal{H}_{0/1}^\dag(k)$. When $\lambda=0$, $\mathcal{H}(k)$ becomes Hermitian. As a result, the following discussion is also applicable for the Hermitian case. In general, Eq.~\ref{E0} with OBC has two different types of nontrivial boundary states, the conventional one that has a Hermitian counterpart, and the non-Hermitian skin modes without Hermitian counterpart. Therefore,  non-Hermitian Hamiltonians have two different types of BBC as shown in Fig.~\ref{F1}. One relates the conventional boundary state to the wavefunction topology of the {\em GBZ Hamiltonian} $\mathcal{H}(\beta_{GBZ})$~\cite{yaoEdgeStatesTopological2018b}. Another relates the non-Hermitian skin modes to the (energy) spectra topology of the {\em Bloch Hamiltonian} $\mathcal{H}(\beta_{BZ})$~\cite{zhangCorrespondenceWindingNumbers2019,Okuma2020}. When the spectra topology is trivial, skin modes do not exist, and GBZ coincides with BZ. As a result, the conventional boundary state can be faithfully predicted by the wavefunction topology of the Bloch Hamiltonian. Actually, the Hermitian Hamiltonian belongs to this case. However, in general, GBZ and BZ are not identical. In this case, if we want to study the boundary states protected by the wavefunction topology, the information of GBZ is necessary. 

\begin{figure}[b]
	\centerline{\includegraphics[height=7cm]{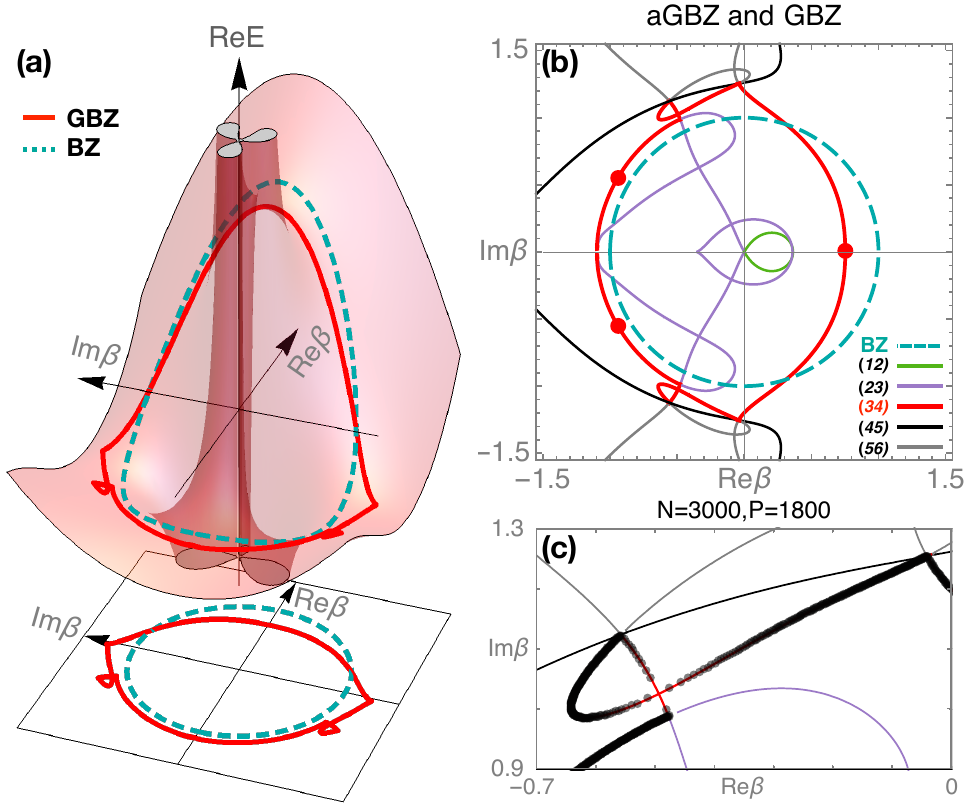}}
	\caption{Non-Hermitian bands, aGBZ, and GBZ of the single band model discussed in the main text. (a) shows the Bloch band and GBZ band can be regarded as different loops on the 2D surface $f(\beta,E)=0$. (b) shows BZ, aGBZ, and GBZ, where different colors represent different root ordering of the analytic arcs, and the red points represent the self-conjugate points satisfying $\beta_p=\beta_{p+1}$. The GBZ is constitute by the $(p,p+1)$ arcs (red one). (c) shows the numerical results with $N=3000$ (lattice size) and $P=1800$ (digit precision). 
		\label{F2}}
\end{figure}

\begin{figure}[t]
	\centerline{\includegraphics[height=6.5cm]{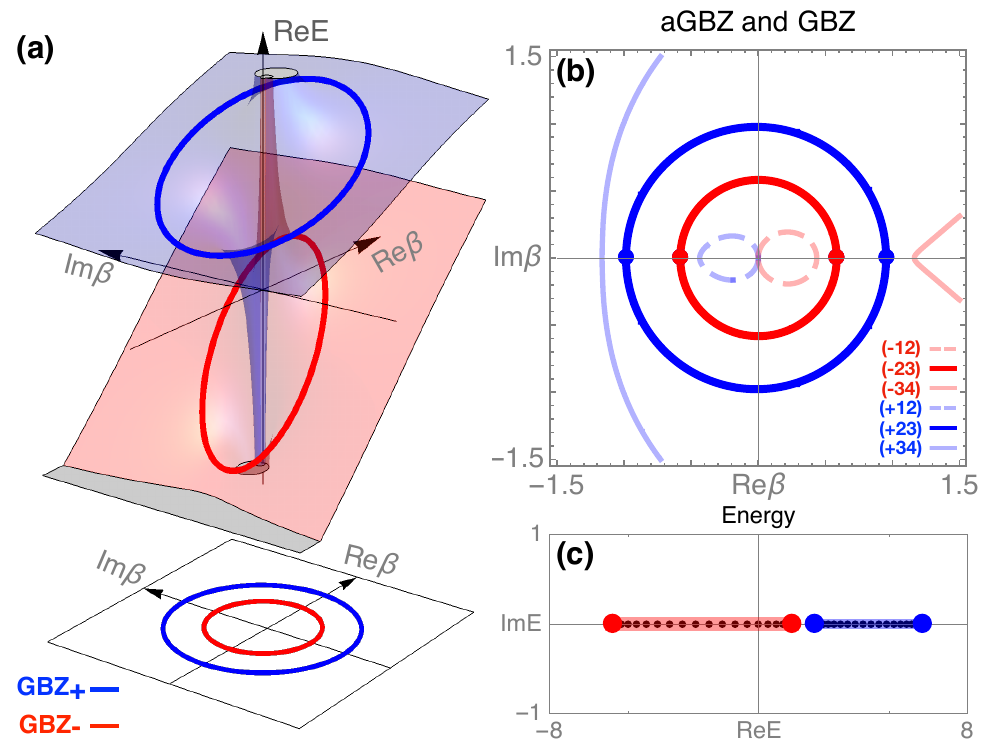}}
	\caption{Non-Hermitian bands, aGBZ, and GBZ of two band model shown in Eq.~\ref{E10}. Different colors in (a) represent different roots in Eq.~\ref{E11}. (b) shows the aGBZ and GBZ. Each analytic arcs on the aGBZ can not only be labeled by the ordering, but also by the band index. The GBZ is constituted by the $(\pm,2,3)$ arcs. (c) shows the numerical eigenvalues (black points) and GBZ spectra (red and blue lines). The red/blue points in (b) and (c) represent the self-conjugate points ($\beta_p=\beta_{p+1}$) of $E_-$/$E_+$ bands, respectively.  
		\label{F3}}
\end{figure}

{\em GBZ and aGBZ.}---In order to characterize the non-Hermitian skin modes, we first extend the crystal momentum from real number to the entire complex plane. Since $\mathcal{H}(k)=\mathcal{H}(k+2\pi n)$, a natural extension of Eq.~\ref{E0} is 
\begin{equation}
\{\mathcal{H}(k),~ k\in\mathbb{R}\}~\rightarrow~\{\mathcal{H}(\beta=e^{ik}),~ k\in\mathbb{C}\}.
\end{equation}
The eigenvalues of $\mathcal{H}(\beta)$ are determined by the following characteristic equation 
\begin{equation}
f(\beta,E)=\det[E-\mathcal{H}(\beta)]=\frac{P(\beta,E)}{\beta^p}=0,
\label{CE}
\end{equation}
where $p$ is the order of the pole of $f(\beta,E)$. For example, in the Hatano-Nelson model $\mathcal{H}(\beta)=\mu+t_1\beta+t_{-1}/\beta$~\cite{PhysRevLett.77.570}, it is obvious that $p=1$ and $P(\beta,E)=-t_1\beta^2+(E-\mu)\beta-t_{-1}$. Geometrically, the characteristic equation Eq.~\ref{CE} defines a 2D (Riemann) surface in the 4D space $(\beta,E)\in\mathbb{C}^2$. According to $f(\beta,E)=\prod_{\mu=1}^n[E-E_{\mu}(\beta)]=0$, each energy band (or root) $E=E_{\mu}(\beta)$ corresponds to a branch of the multivalued function. When the boundary condition is fixed to PBC or OBC, the corresponding Bloch band ($\{E_\mu(\beta_{BZ,\mu}),\mu=1,...,n\}$) or GBZ band ($\{E_\mu(\beta_{GBZ,\mu}),\mu=1,...,n\}$) become a set of closed loops on the Riemann surface. As shown in Fig.~\ref{F2} (a) and Fig.~\ref{F3} (a), the GBZ is the projection of the GBZ band on the complex $\beta$-plane.

The aGBZ is defined by the projection of the following two equations on the complex $\beta$-plane, 
\begin{equation}\label{E4}
f(\beta,E)=f(\beta e^{i\theta},E)=0,\quad \theta\in\mathbb{R}.
\end{equation}
The mathematical meaning of aGBZ is that for a given point $\beta_0$ on it with $f(\beta_0,E_0)=0$, there must exist a conjugate point $\tilde{\beta}_0=\beta_0 e^{i\theta_0}$ on it satisfying $f(\tilde{\beta}_0,E_0)=0$~\cite{FN1}. Therefore, one can define the {\em root ordering} of $\beta_0\in\beta_{aGBZ}$ via the following procedure: (i) solve $f(\beta,E_0)=0$; (ii) order the roots by the absolute value; (iii) identify the ordering of two roots that have the same absolute value as $|\beta_0|$. For example, if $|\beta_0|=|\beta_m(E_0)|=|\beta_{m+1}(E_0)|$, then, $(m,m+1)$ is the root ordering of $\beta_0$. This root ordering will be used to pick GBZ from the aGBZ. Since there exist five variables $(\Re \beta,\Im \beta,\Re E,\Im E,\theta)$ and four constraint equations $\Re f=\Im f=\Re f^\theta=\Im f^\theta=0$, where $f^\theta:=f(\beta e^{i\theta},E)$, the solution of Eq.~\ref{E4} is 1D curve in the 5D space. When the additional degrees, $\theta$ and $E$, are eliminated, it can be shown that the constraint equation of the aGBZ is an algebraic equation of $\Re\beta$ and $\Im\beta$,
\begin{equation}
F_{aGBZ}(\Re \beta,\Im \beta)=\sum_{i,j}c_{ij}(\Re\beta)^i(\Im\beta)^j=0.
\label{E5}
\end{equation}
In the Supplemental Material (SM), we show how to prove Eq.~\ref{E5} and calculate the coefficients $c_{ij}$ by using the concept of resultant~\cite{Resultant2019,woodyDepartmentMathematicsComputer,jansonRESULTANTDISCRIMINANTPOLYNOMIALS,SM2}. The solid lines in Fig.~\ref{F2} (b) with different colors show an example of aGBZ  of the following model  $\mathcal{H}(\beta)=-1/6-1/(2\beta^3)+8/(5\beta^2)+10/(3\beta)+4\beta+2\beta^2+\beta^3$.  Obviously, the aGBZ is constituted by a set of analytic arcs joined by the self-intersection points. 

Now we show how to obtain the GBZ from aGBZ. Notice that any analytic arc on the aGBZ can be labeled by a common root ordering, as shown in Fig.~\ref{F2} (b) with different colors~\cite{FN2}. For the single-band models, the GBZ is constituted by all the arcs labeled by $(p,p+1)$~\cite{zhangCorrespondenceWindingNumbers2019,Okuma2020}, e.g., $(3,4)$ in our example. As shown in Fig.~\ref{F2} (c) and SM, our analytical result is consistent with the numerical results with $N=3000$ (lattice size) and $P=1800$ (digit precision). However, according to the numerical result, we do not know whether there exist self-intersection points on the GBZ~\cite{FN3}. We note that under the current $N$ and $P$, the calculation time is 11 days. If we continue to improve the lattice size and numerical accuracy, the calculation time will become unacceptable. We further note that if the calculation precision is not so high ($P=1800$), the numerical result for $N=3000$ may become  incorrect~\cite{SM2}. This is the central difficulty of the numerical calculation: the numerical diagonalization error of non-Hermitian Hamiltonians is highly sensitive to the matrix size and sometimes may lead to incorrect simulations~\cite{SM2,colbrookHowComputeSpectra2019b,bottcher2005,REICHEL1992153}. Our analytical method overcomes this difficulty and can be further used to verify the accuracy of numerical calculations. On the GBZ, there exists a set of {\em self-conjugate points} satisfying $\tilde{\beta}_p=\beta_{p+1}$, as shown in Fig.~\ref{F2} (b) with red points. A statement about the self-conjugate points is that any analytic arc containing them must form the GBZ. In summary, the aGBZ is a minimal analytic element containing all the information of GBZ and the GBZ is in general a subset of aGBZ. 

Generalizing the discussion to the multi-band system, we will show that the sub-GBZs for each band can be distinct. Consider the following two-band example, 
\begin{equation}
\mathcal{H}(\beta)=\left(\begin{array}{cc}{t_0+t_{-1}/\beta+t_1\beta} & {c} \\ {c} & {w_0+w_{-1}/\beta+w_1\beta}\end{array}\right). 
\label{E10}
\end{equation}
with $t_0=4,t_1=t_{-1}=1,w_0=-2,w_1=3,w_{-1}=1,c=-1$. The eigenvalues of the Hamiltonian are 
\begin{equation}
E_{\pm}(\beta)=h_0(\beta)\pm\sqrt{c^2+h_z^2(\beta)},
\label{E11}
\end{equation}
where $h_{0/z}(\beta)=[h_1(\beta)\pm h_2(\beta)]/2$,  $h_1(\beta)=t_0+t_{-1}/\beta+t_1\beta$, $h_2(\beta)=w_0+w_{-1}/\beta+w_1\beta$. As shown in Fig.~\ref{F3} (a), the red and blue surfaces show the real parts of $E_+(\beta)$ and $E_-(\beta)$, respectively. When the OBC is chosen, $E_{\pm}(\beta_{GBZ,\pm})$ (red/blue solid lines) define two closed loop on the branches $E_{\pm}(\beta)$, respectively. As shown in Fig.~\ref{F3} (a), their projections on the complex plane are the multi-band GBZ, which is constituted by two distinct sub-GBZs, $\beta_{GBZ,+}$ and $\beta_{GBZ,-}$. 

If the multi-band Hamiltonian is not block diagonal and has no additional symmetry, the GBZ is also constituted by all the arcs labeled by $(p,p+1)$~\cite{yaoEdgeStatesTopological2018b,yokomizoNonBlochBandTheory2019a,zhangCorrespondenceWindingNumbers2019,2020arXiv200302219Y}, e.g., $(2,3)$ in our example with $c\neq 0$. However, if the Hamiltonian is block diagonal, e.g., $c=0$ in Eq.~\ref{E10}, then, the GBZ is the union of the ones belonging to each non-block diagonal part, namely, $\beta_{GBZ}=\beta_{GBZ,1}\cup \beta_{GBZ,2}$ for $c=0$ in Eq.~\ref{E10}~\cite{FN4}. We now extract the band information from aGBZ. From the aGBZ (dashed and solid lines) shown in Fig.~\ref{F3} (b), for any point $\beta_0$ on the analytic arc, Eq.~\ref{E11} maps $\beta_0$ to $E_+(\beta_0)$ and $E_-(\beta_0)$. By solving $f(\beta,E_{\pm}(\beta_0))=0$ and ordering the roots by the absolute values, one can check which one satisfies the aGBZ condition, that is, there exist two roots having the same absolute values as $|\beta_0|$. Therefore, all the analytic arcs can be further labeled by the band index. For example, the blue/red lines in Fig.~\ref{F3} (b) belong to $E_{\pm}$ band, respectively. In our example, only the arcs with labeling $(\pm,p,p+1)$ constitute the GBZ. Using Eq.~\ref{E11} to map  $\beta_{GBZ,\pm}$ to $E_\pm(\beta_{GBZ,\pm})$, one can obtain the GBZ spectra shown in Fig.~\ref{F3} (c) with blue and red lines, which matches the numerical results (black dots)~\cite{FN5}. The self-conjugate points (red and blue points) in Fig.~\ref{F3} (b) and (c) correspond to the end points of the energy spectra. We finally note that each band, $E_\mu(\beta)$, can only map its own sub-GBZ, $\beta_{GBZ,\mu}$. This fact has a geometrical interpretation: each band dispersion is only defined on each branch of Eq~\ref{CE}.

\begin{figure}[t]
	\centerline{\includegraphics[height=7cm]{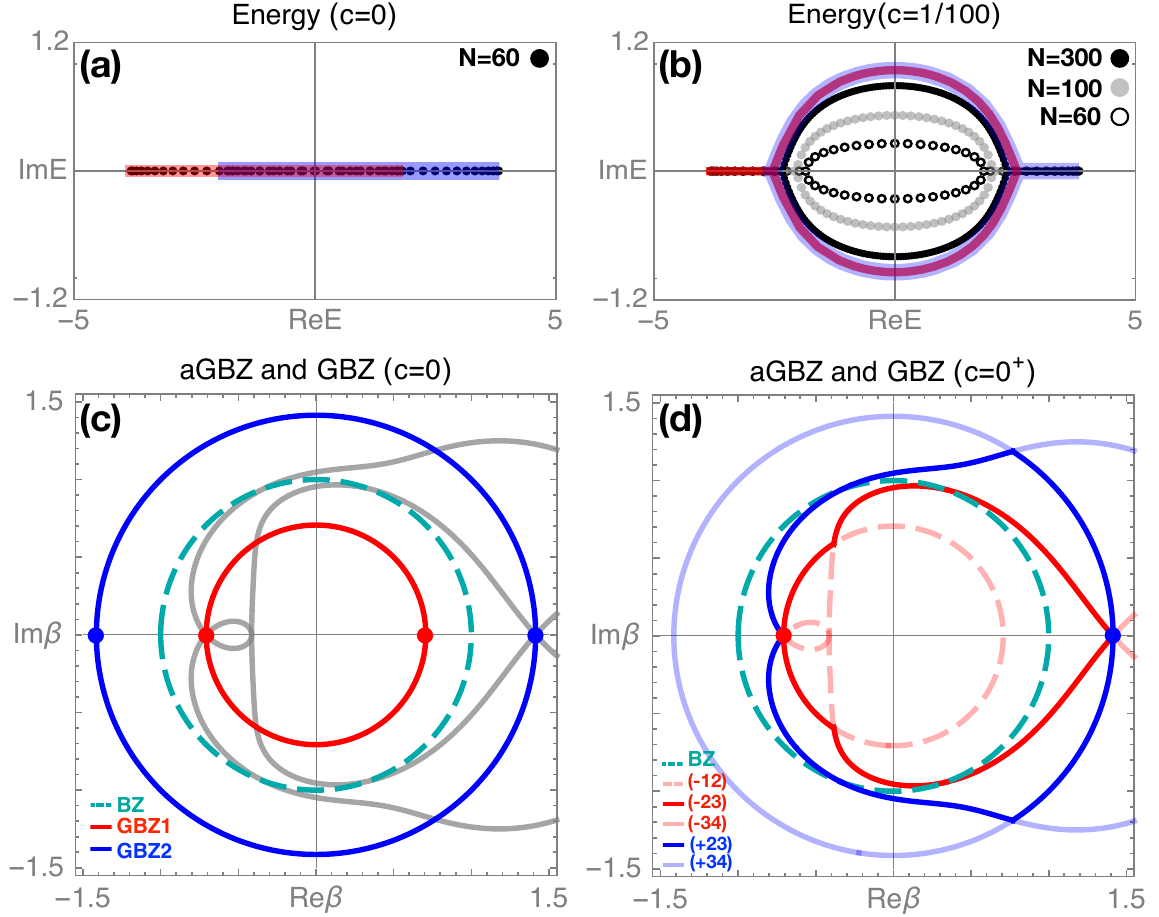}}
	\caption{Perturbation-failure (or critical skin) effect. The  non-perturbative behavior of the OBC spectrum between $c=0$ (dots in (a)) and $c=1/100$ (dots in (b)) of Eq.~\ref{E10} can be understood by the discontinuity of GBZ (red and blue opaque solid lines) in (c) and (d), namely, $\beta_{GBZ}(c=0)\neq\beta_{GBZ}(c=0^+)$. The solid lines in (a) and (b) represent the analytic GBZ spectrum $E_{GBZ}(c=0)$ and $E_{GBZ}(c=0^+)$, respectively. 
		\label{F4}}
\end{figure}

{\em Application I: perturbation-failure.}---We now show some applications of the aGBZ theory. The first one is the perturbation-failure (or critical skin) effect in non-Hermitian band theory~\cite{PhysRevLett.123.097701,LinhuLi2020}. As shown in Fig.~\ref{F4} (a) and (b), when we choose $t_0=1,t_1=1,t_{-1}=2,w_0=-1,w_1=3,w_{-1}=1$ in Eq.~\ref{E10}, the OBC spectrum (dots) of $c=0$ and $c=1/100$ exhibits a non-perturbative behavior~\cite{FN6}. With the increasing of lattice size $N$, the non-perturbative effect becomes stronger. The aGBZ theory not only provides an analytical method to understand this phenomenon, but also can strictly prove the discontinuity of the energy spectrum evolution at $c=0$ under the thermodynamic limit~\cite{FN7}. As shown in Fig.~\ref{F4} (c) and (d), when $c=0$ and $c=0^+$ (right-hand limit), the aGBZ of Eq.~\ref{E10} are the same, namely, $\beta_{aGBZ}(c=0)=\beta_{aGBZ}(c=0^+)$. However, when $c$ changes from zero to nonzero, the GBZ condition is changed. To be more precise, when $c=0$, Eq.~\ref{E10} is diagonal and the characteristic equation $f(\beta,E)=[E-h_1(\beta)][E-h_2(\beta)]$ is {\em reducible}. The asymptotic solutions are determined by the two separated irreducible polynomials $E-h_1(\beta)$ and $E-h_2(\beta)$, which result two independent sub-GBZs,  $\beta_{GBZ,1}=\sqrt{t_{-1}/t_{1}}e^{ik}$ and $\beta_{GBZ,2}=\sqrt{w_{-1}/w_{1}}e^{ik}$, as shown in Fig.~\ref{F4} (c). However, when $c\rightarrow 0^+$, these two bands will couple together and the GBZ is determined by the {\em irreducible polynomial}  $f(\beta,E)=[E-h_1(\beta)][E-h_2(\beta)]-c^2$. As a result, only the $(\pm,2,3)$ arcs on the aGBZ constitute the GBZ, as shown in Fig.~\ref{F4} (d). Comparing (c) and (d), it is obvious $\beta_{GBZ}(c=0)\neq\beta_{GBZ}(c=0^+)$, which implies $E_{GBZ}(c=0)\neq E_{GBZ}(c=0^+)$ as shown by the solid   lines in Fig.~\ref{F4} (a) and (b), respectively.

\begin{figure}[t]
	\centerline{\includegraphics[height=7cm]{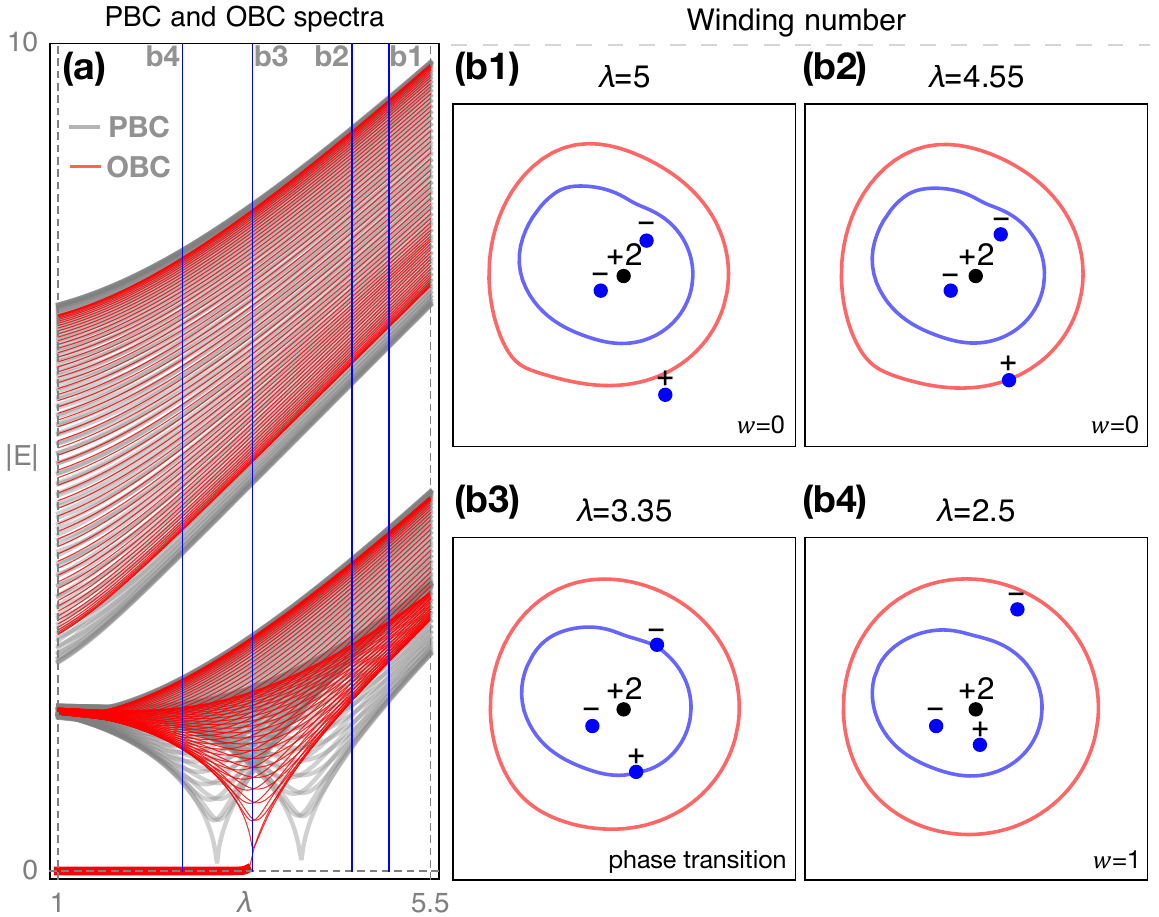}}
	\caption{Wavefunction winding number and non-degenerate sub-GBZs of Eq.~\ref{E12}. (a) shows the PBC/OBC spectrum $|E|$ as a function of $\lambda$. (b) shows the evolution of GBZ, topological charge, and winding number. Red and blue lines represent two distinct sub-GBZs. The total winding number equals one half of the charge summation of the black dot and the blue dots inside the blue sub-GBZ. 
		\label{F5}}
\end{figure}

{\em Application II: wavefunction winding number.}---The second application of the aGBZ theory is the BBC in the case where each band has its respective, distinct sub-GBZ. Consider the following {\em four-band} model preserving sub-lattice symmetry~\cite{PhysRevX.9.041015}, 
\begin{equation}
\mathcal{H}(\beta)=\left(\begin{array}{cc}
0 & R_{+}(\beta) \\
R_{-}(\beta) & 0
\end{array}\right),
\label{E12}
\end{equation}
where $R_{\pm}(\beta)=\lambda+(t_\pm+t_1\beta^{\pm 1})\sigma_\pm+t_2\beta^{\pm 1}\sigma_\mp$ and $t_1=2, t_2=2i, t_{\pm}=5\pm2$. Since the Hamiltonian has sub-lattice symmetry, the eigenvalues come in pairs, e.g., $(E,-E)$. As a result, the sub-GBZs of $E_{\mu}(\beta)$ and $-E_{\mu}(\beta)$ must be degenerate~\cite{FN8}. Fig.~\ref{F5} (a) shows the differences between OBC/PBC spectrum ($|E|$) as $\lambda$ evolves. In order to characterize the emergence of topological zero modes in (a), we need to define the (wavefunction) winding number of the GBZ Hamiltonian $\mathcal{H}(\beta_{GBZ})$. However, due to the existence of multiple sub-GBZs shown in Fig.~\ref{F5} (b), the definition based on the $Q$ matrix~\cite{yaoEdgeStatesTopological2018b,yokomizoNonBlochBandTheory2019a} can not be extended directly~\cite{FN9}. We note that once the root of $\det[\mathcal{H}(\beta)]=0$ passes through the GBZ, it {\em may} correspond to a topological phase transition. Therefore, it can be regarded as a topological charge. According to $\det[\mathcal{H}(\beta)]=\det[R_+(\beta)]\det[R_-(\beta)]=E_1^2(\beta)E_2^2(\beta)=0$, the zeros can be labeled by $R$ index $\pm$, which determine the sign of the charge, and band index $\mu=1,2$, which is related to the sub-GBZs. When the zeros belonging to the first band ($E_1(\beta)=0$) cross the sub-GBZ of second band ($\beta_{GBZ,2}$), as shown in Fig.~\ref{F5} (b2) where the colors represent the band index, there is no gap closing and phase transition. This inspires us to write down the following conjectured formula~\cite{SM2} 
\begin{equation}
w=\frac{1}{2}(w_+-w_-),\quad w_\pm=-P_{\pm}+\sum_{\mu=1}^mZ_{\pm,\mu}
\end{equation}
where $Z_{\pm,\mu}$ are the number of zeros not only satisfying $\det[R_\pm(\beta)]=E_\mu(\beta)=0$ but also being inside the sub-GBZ $\beta_{GBZ,\mu}$, and $P_{\pm}$ are the orders of the pole of $\det[R_\pm(\beta)]$. As shown in Fig.~\ref{F5} (b), we plot the GBZ and the topological charges for different values of $\lambda$, where the black dots represent the charge of pole, namely, $P_+=0$ and $P_-=2$; the blue dots with charge $\pm$ represent the zeros belonging to the blue sub-GBZ band and satisfying $\det[R_\pm(\beta)]=0$. Since there are no zeros belonging to the red sub-GBZ band under the parameters shown in (b), the total winding number equals one half of the charge summation of the black dot and the blue dots inside the blue sub-GBZ. This result is consistent with Fig.~\ref{F5} (a). 

{\em Discussions and conclusions.}---In summary, we have provided an analytical method to calculate the GBZ, which acts as the role of the exact solution of non-Hermitian OBC Hamiltonians in the thermodynamic limit. Compared with the previous numerical methods, our work reduces the problem to the task of calculating the resultant and solving algebraic equations, the process of which is faster and error-free. 

\begin{acknowledgments}
The work is supported by the Ministry of Science and Technology of China 973 program (No.~2017YFA0303100), National Science Foundation of China (Grant No.  NSFC-11888101, 1190020, 11534014, 11334012), and the Strategic Priority Research Program of CAS (Grant No.XDB07000000). 	
\end{acknowledgments}

\bibliography{aGBZ}

\begin{thebibliography}{73}
\bibitem [{FN1()}]{FN1}%
  \BibitemOpen
  \href@noop {} {\bibinfo  {journal} {In general, $\beta_0$ and
  $\tilde{\beta}_0$ are different points on the aGBZ}\ }\BibitemShut {NoStop}%
\bibitem [{FN2()}]{FN2}%
  \BibitemOpen
\bibfield  {journal} {  }\href@noop {} {\bibinfo  {journal} {The reason is that
  for any point on the analytic arcs, the root ordering can only be changed at
  the self-intersection points}\ }\BibitemShut {NoStop}%
\bibitem [{FN3()}]{FN3}%
  \BibitemOpen
\bibfield  {journal} {  }\href@noop {} {\bibinfo  {journal} {We note that this
  kind of self-intersection point is a new type of singularity point, which has
  no counterpart in Hermitian systems and is unique to non-Hermitian systems.
  The details have been discussed in the SM}\ }\BibitemShut {NoStop}%
\bibitem [{FN4()}]{FN4}%
  \BibitemOpen
\bibfield  {journal} {  }\href@noop {} {\bibinfo  {journal} {It can be shown
  that $\beta_{GBZ,1}=\sqrt{t_{-1}/t_{1}}e^{ik}$ and
  $\beta_{GBZ,2}=\sqrt{w_{-1}/w_{1}}e^{ik}$ for the parameters we chosen, i.e.,
  $t_0=4,t_1=t_{-1}=1,w_0=-2,w_1=3,w_{-1}=1$.}\ }\BibitemShut {NoStop}%
\bibitem [{FN5()}]{FN5}%
  \BibitemOpen
\bibfield  {journal} {  }\href@noop {} {\bibinfo  {journal} {We note that
  $\beta_{GBZ,+}$ can only be mapped by $E_{+}(\beta)$ and the corresponding
  GBZ spectra for $E_+$ band becomes $E_+(\beta_{GBZ,+})$. While the GBZ
  spectra for $E_-$ band becomes $E_-(\beta_{GBZ,-})$}\ }\BibitemShut {NoStop}%
\bibitem [{FN6()}]{FN6}%
  \BibitemOpen
  \href@noop {} {\bibinfo  {journal} {This means the spectra change is lager
  than the perturbation strength}\ }\BibitemShut {NoStop}%
\bibitem [{FN7()}]{FN7}%
  \BibitemOpen
\bibfield  {journal} {  }\href@noop {} {\bibinfo  {journal} {We note that the
  GBZ and the corresponding GBZ spectra $E_{\mu}(\beta_{GBZ,\mu})$ are defined
  in the thermodynamic limit}\ }\BibitemShut {NoStop}%
\bibitem [{FN9()}]{FN9}%
  \BibitemOpen
\bibfield  {journal} {  }\href@noop {} {\bibinfo  {journal} {The reason is that
  the Q matrix is defined on all occupied bands, it is not clear which sub-GBZs
  should be chosen as the integration path. In the SM, we provide a detailed
  discussion}\ }\BibitemShut {NoStop}%
\end{thebibliography}
\bibliographystyle{apsrev4-1}

\onecolumngrid
\newpage


\section*{Supplementary material}

# Supplemental Material for
# "Non-Hermitian bulk-boundary correspondence and auxiliary generalized Brillouin zone theory"

Zhesen Yang[1,2,][*], Kai Zhang[1,2,][*], Chen Fang[1,][†] and Jiangping Hu[1,2,3,4,][‡]

[1]*Beijing National Laboratory for Condensed Matter Physics,*
*and Institute of Physics, Chinese Academy of Sciences, Beijing 100190, China*
[2]*University of Chinese Academy of Sciences, Beijing 100049, China*
[3]*Kavli Institute for Theoretical Sciences and CAS Center for Excellence in Topological Quantum Computation,*
*University of Chinese Academy of Sciences, Beijing 100190, China and*
[4]*South Bay Interdisciplinary Science Center, Dongguan, Guangdong Province, China*

## Contents

## I. DIFFICULTIES IN NUMERICAL CALCULATION OF GENERALIZED BRILLOUIN ZONE

In the main text, we have mentioned that the numerical calculation method of generalized Brillouin zone (GBZ) is not only time-consuming but also unreliable due to the numerical diagonalization errors that are often sensitive to the matrix dimension and calculation accuracy. In this section, we will explain this point.

### A. Review of numerical calculation of generalized Brillouin zone

We first review the numerical calculation method of GBZ proposed in Ref. [1, 2], in which the procedure can be summarized as follows:

---

[*]These two authors contributed equally
[†]Corresponding author: cfang@iphy.ac.cn
[‡]Corresponding author: jphu@iphy.ac.cn

- Write down the non-Bloch Hamiltonian $\mathcal{H}(\beta)$, which can be obtained from the Bloch Hamiltonian $\mathcal{H}(k)$ by replacing $e^{ik} \to \beta$, where $\beta \in \mathbb{C}$. Then, calculate the characteristic equation of the non-Bloch Hamiltonian

$$f(\beta, E) = \det[E - \mathcal{H}(\beta)] = \frac{P(\beta, E)}{\beta^p}, \tag{1}$$

where $p$ is the order of the pole of $f(\beta, E)$ and $P(\beta, E)$ is an algebraic polynomial in variables $\beta$ and $E$. We assume $P(\beta, E)$ has degree $p + s$ in $\beta$.

- Diagonal the same Hamiltonian with open boundary condition $H_N$, whose lattice size is $N$, and obtain the corresponding eigenvalues

$$S_N := \{E_i \in \mathbb{C} : E_i \text{ is the eigenvalue of } H_N\}. \tag{2}$$

- For any given $E_i \in S_N$, solve the characteristic equation of the non-Bloch Hamiltonian

$$f(\beta, E_i) = 0, \tag{3}$$

and order the roots by the absolute values, e.g.,

$$|\beta_1(E_i)| \le |\beta_2(E_i)| \le ... \le |\beta_{p+s-1}(E_i)| \le |\beta_{p+s}(E_i)|. \tag{4}$$

- If the Hamiltonian is not block diagonal and does not have additional symmetries, the GBZ is constituted by the $p$th and $(p+1)$th roots

$$\beta_{GBZ}(N) = \{\beta_p(E_i) \in \mathbb{C} : E_i \in S_N\} \cup \{\beta_{p+1}(E_i) \in \mathbb{C} : E_i \in S_N\}. \tag{5}$$

Here $N$ represents the lattice size.

The above procedure can be applied to both single-band and multi-band Hamiltonians. Based on the above procedure, the numerical diagonalized eigenvalues of the open boundary Hamiltonian $H_N$ are required. However, as will be discussed in the following subsection, these eigenvalues are often sensitive to the matrix dimension and calculation accuracy. We note that if the Hamiltonian has additional symmetry, e.g., spinful anomalous time-reversal symmetry, the GBZ condition Eq. 4 can be changed [3, 4].

### B. Difficulties

As mentioned above, the numerical calculation of GBZ requires the information of the open boundary energy spectrum. If the diagonalized eigenvalues of $H_N$ have large numerical errors, the calculated GBZ will become incorrect.

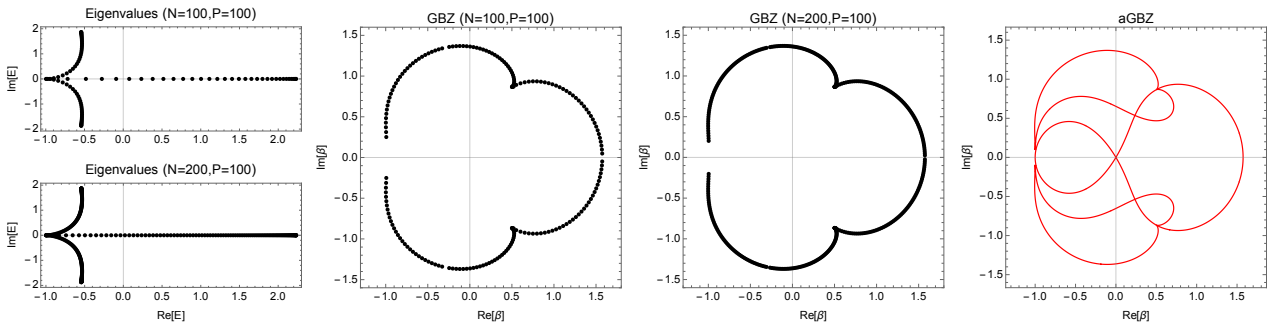

FIG. 1: Numerically calculated eigenvalues, the corresponding GBZs, and auxiliary GBZ of Eq. 6. Here $N$ and $P$ represent the lattice size and digit precision, respectively.

We first use an example to show that, sometimes, the information of the numerical calculated GBZ may be incomplete when the lattice size $N$ is not so large. As shown in Fig. 1, the numerical calculated eigenvalues and the corresponding GBZs of the following model are plotted

$$\mathcal{H}(\beta) = \beta + 1/\beta^2 + 1/\beta^3. \tag{6}$$

One can notice that because the eigenvalues near $E = -1$ on the real axis are very loose, the corresponding GBZ around $\beta = -1$ are unclear, and seems disconnected. If we want to accurately describe the behavior of GBZ around $\beta = -1$, the lattice size $N$ should be very large. However, as will be discussed in the following example, a large $N$ requires a large calculation accuracy $P$, and accordingly, the calculation time $T$ will become unacceptable. As shown in Fig. 1, our analytical results (red lines) do not have this problem. Notice that we only plot the aGBZ in Fig. 1. The relation between aGBZ and GBZ has been discussed in the main text.

Now we use another example to show that the numerical diagonalization error of the non Hermitian matrix is often sensitive to the matrix dimension and calculation accuracy. In Fig. 2, we plot the numerical eigenvalues (red points) of the following non-Bloch Hamiltonian with open boundary condition

$$\mathcal{H}(\beta) = 5/\beta^5 + 4/\beta^4 + 3/\beta^3 + 2/\beta^2 + 1/\beta - \beta, \tag{7}$$

under different choices of $N$ (lattice size) and $P$ (digit precision). We find that the numerical energy spectrum (from left to right in each row) significant varies as the lattice size $N$ increases under fixed digit precision $P$. Therefore, in order to obtain the reliable eigenvalues in the large $N$ case, the numerical accuracy must be increased. However, the corresponding calculation time $T$ will become longer. More importantly, for any given lattice size $N$, the minimal calculation accuracy $P$ is unknown. This is an intrinsic issue in the numerical calculation of non-Hermitian matrices, and sometimes, may lead to incorrect simulations. Actually, this point has been discussed in a recent work, i.e., Ref. [5], in which, the authors mainly studied how to control errors in the numerical diagonalization. However, as discussed in the main text, our analytical method overcomes this difficulty, and can even be applied to verify the accuracy of numerical calculations. To be more precise, we can first use our method to calculate the GBZ $\beta_{GBZ}$, and then, the analytical spectra (or GBZ spectra) $E_{GBZ}$ can be obtained through the Hamiltonian Eq. 7, i.e., $E_{GBZ} = \mathcal{H}(\beta_{GBZ})$. In Fig. 2, the analytical spectra is plotted with blue line, which is consistent with some of the numerical results.

## II. CALCULATION OF AUXILIARY GENERALIZED BRILLOUIN ZONE

In the main text, we have mentioned that the aGBZ is an algebraic polynomial of $\mathrm{Re}\,\beta$ and $\mathrm{Im}\,\beta$, and can be calculated by the resultant method of two polynomials. Here we briefly review the definition of resultant and show how to calculate them. Finally, we will use an example to help the readers to have a better understanding of the resultant.

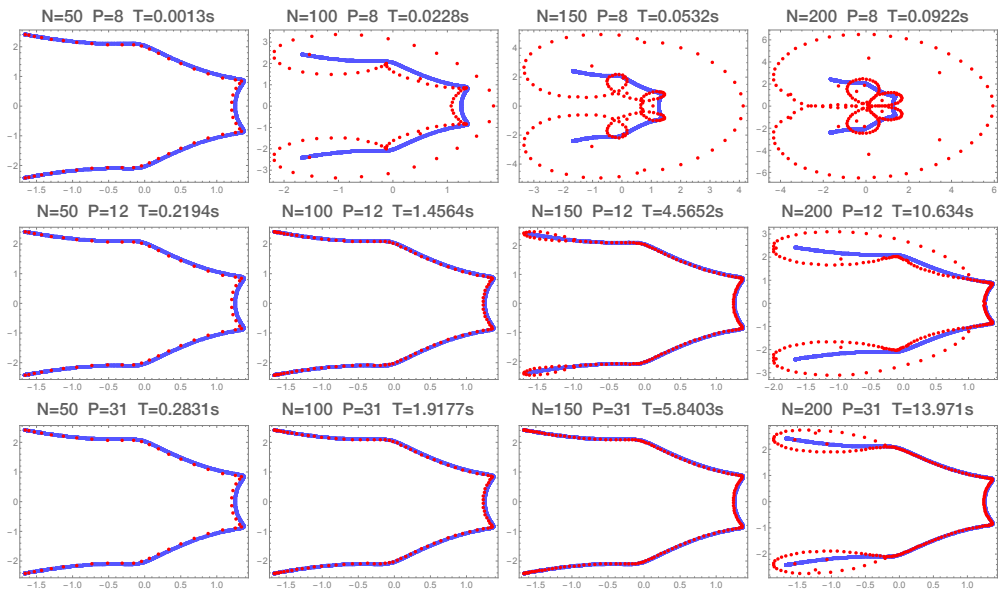

FIG. 2: The comparison between the numerical diagonalization eigenvalues (red points) and the GBZ spectra calculated by our analytical method (blue lines). Here $N$, $P$ and $T$ represent the lattice size, digit precision and calculation time, respectively.

## A. Resultant of two polynomials

**Definition A.1** (Polynomial). A polynomial $f(x) \in F[x]$ is defined as

$$f(x) = \prod_{i=1}^{n}(x - \xi_i) = a_n x^n + a_{n-1}x^{n-1} + ... + a_1 x + a_0, \quad a_n \neq 0 \tag{8}$$

where each coefficient $a_i$ belongs to the field $F$ and each root $\xi_i$ belongs to the extension of $F$. For example, if $a_n, ..., a_0$ are real numbers, $\xi_1, ..., \xi_n$ are complex numbers.

**Definition A.2** (Resultant). Given two polynomials $f(x) = a_n x^n + ... + a_0$, $g(x) = b_m x^m + ... + b_0 \in F[x]$, their resultant relative to the variable $x$ is a polynomial over the field of coefficients of $f(x)$ and $g(x)$, and is defined as

$$R_x(f, g) = a_n^m b_m^n \prod_{i,j}(\xi_i - \eta_j), \tag{9}$$

where $f(\xi_i) = 0$ for $1 \leq i \leq n$ and $g(\eta_j) = 0$ for $1 \leq j \leq m$.

**Theorem A.3**. Let $f(x) = a_n x^n + ... + a_0$, $g(x) = b_m x^m + ... + b_0 \in F[x]$,

1. *Suppose that $f$ has $n$ roots $\xi_1, ..., \xi_n$ in some extension of $F$. Then*

$$R_x(f, g) = a_n^m \prod_{i=1}^{n} g(\xi_i). \tag{10}$$

2. *Suppose that $g$ has $m$ roots $\eta_1, ..., \eta_m$ in some extension of $F$. Then*

$$R_x(f, g) = (-1)^{mn} b_m^n \prod_{j=1}^{m} f(\eta_j). \tag{11}$$

The proof can be found in Ref [6, 7].

**Theorem A.4**. *Let $f$ and $g$ be two non-zero polynomials with coefficients in a field $F$. Then $f$ and $g$ have a common root in some extension of $F$ if and only if their resultant $R_x(f, g)$ is equal to zero.*

*Proof*: Suppose $\gamma$ is their common root, $R_x(f, g) \propto (\gamma - \gamma) = 0$. Conversely, If $R_x(f, g) = 0$, at least one of the factors of $R_x(f, g)$ must be zero, say $\xi_i - \eta_j = 0$, then, $\xi_i = \eta_j$ is their common root.

From Theorem A.4, the resultant can be applied to make sure whether two polynomials share a common root. However, from Definition A.2, the calculation of the resultant requires to know the roots of each polynomial. The following theorem enables us to calculate the resultant directly according to the coefficients of $f$ and $g$.

**Definition A.5**. The Sylvester matrix of two polynomials $f(x) = a_n x^n + ... + a_0$, $g(x) = b_m x^m + ... + b_0 \in F[x]$ is defined by

$$\text{Syl}(f, g) = \begin{pmatrix} a_n & a_{n-1} & a_{n-2} & \ldots & 0 & 0 & 0 \\ 0 & a_n & a_{n-1} & \ldots & 0 & 0 & 0 \\ \vdots & \vdots & \vdots & & \vdots & \vdots & \vdots \\ 0 & 0 & 0 & \ldots & a_1 & a_0 & 0 \\ 0 & 0 & 0 & \cdots & a_2 & a_1 & a_0 \\ b_m & b_{m-1} & b_{m-2} & \ldots & 0 & 0 & 0 \\ 0 & b_m & b_{m-1} & \cdots & 0 & 0 & 0 \\ \vdots & \vdots & \vdots & & \vdots & \vdots & \vdots \\ 0 & 0 & 0 & \cdots & b_1 & b_0 & 0 \\ 0 & 0 & 0 & \cdots & b_2 & b_1 & b_0 \end{pmatrix}, \tag{12}$$

where $a_n, ..., a_0$ are the coefficients of $f$ and $b_m, ..., b_0$ are the coefficients of $g$.

**Theorem A.6**. *The resultant of two polynomials $f, g$ equals to the determinant of their Sylvester matrix, namely*

$$R_x(f, g) = \det[\text{Syl}(f, g)] \tag{13}$$

For example, if $n = 3, m = 2$,

$$R_x(f,g) = \det \begin{pmatrix} a_3 & a_2 & a_1 & a_0 & 0 \\ 0 & a_3 & a_2 & a_1 & a_0 \\ b_2 & b_1 & b_0 & 0 & 0 \\ 0 & b_2 & b_1 & b_0 & 0 \\ 0 & 0 & b_2 & b_1 & b_0 \end{pmatrix}. \tag{14}$$

The proof of the theorem can be found in Ref [6, 7]. When $f$ and $g$ have multiple variables, the resultant provides a very simple elimination method. Here we take a simple example to illustrate this point.

**Example A.7.** Consider

$$f(x,y) = (x-y)(x-4y), \qquad g(x,y) = (x-2y)(x-3y), \tag{15}$$

where $x$ and $y$ are two complex variables. We want to eliminate $x$ and obtain a function of $y$ from $f(x,y) = 0$ and $g(x,y) = 0$. The resultant helps to do this because the effect of resultant is to eliminate one variable, and obtain the polynomial only about other variables. According to $f(x,y) = 0$, we have $x_{f,1} = y$ and $x_{f,2} = 4y$. According to $g(x,y) = 0$, we have $x_{g,1} = 2y$ and $x_{g,2} = 3y$. According to definition A.2, if these two polynomials has the common roots of $x$, which means $f(x,y) = 0$ and $g(x,y) = 0$ have solutions, then the resultant is always zero

$$r(y) := R_x[f(x,y), g(x,y)] = \Pi_{i,j}[x_{f,i}(y) - x_{g,j}(y)] = (y-2y)(y-3y)(4y-2y)(4y-3y) = 4y^4 = 0, \tag{16}$$

where the subscript $x$ of $R_x$ represents the variable that is eliminated in the resultant, and we will take this notation in the following discussion. Finally, $r(y) = 0$ is the polynomial equation obtained by eliminating $x$ from $f(x,y) = 0$ and $g(x,y) = 0$.

**Example A.8.** Now we use Theorem A.6 to calculate the resultant. Consider

$$f(x) = x^2 + ax + b, \qquad g(x) = x^2 + cx + d, \tag{17}$$

where $a, , b, c, d$ can be functions of $y$. According to Sylvester matrix, the resultant of $f$ and $g$ is

$$R_x(f,g) = \det \begin{pmatrix} 1 & a & b & 0 \\ 0 & 1 & a & b \\ 1 & c & d & 0 \\ 0 & 1 & c & d \end{pmatrix} = a^2 d - abc - acd + b^2 + bc^2 - 2bd + d^2. \tag{18}$$

Thus $R_x(f,g) = 0$ is the polynomial equation obtained by eliminating $x$ from $f(x) = 0$ and $g(x) = 0$. For more complex polynomials, the **Resultant** command in *Mathematica* can help us to calculate their resultant.

We finally noted that the resultant $R_x(f,g)$ is only defined for two algebraic functions $f(x)$ and $g(x)$. According to Eq. 13, $R_x(f,g)$ must also be an algebraic function that is constituted by the coefficients of $f$ and $g$.

## B. Auxiliary GBZ

In this section, we use the resultant method to calculate the aGBZ and prove that Eq. 5 in the main text is an algebraic polynomial of $\mathrm{Re}\,\beta$ and $\mathrm{Im}\,\beta$. Starting from the following equation,

$$f(\beta, E) = f(\beta e^{i\theta}, E) = 0, \qquad \theta \in \mathbb{R}, \tag{19}$$

our strategy is to eliminate E first, and then $\theta$. Since $f(\beta, E)$ and $f(\beta e^{i\theta}, E)$ are algebraic functions of $E$, their resultant relative to $E$ can be calculated directly from Eq. 13, which is labeled by $G(\beta, \theta)$

$$G(\beta, \theta) := R_E[f(\beta, E), f(\beta e^{i\theta}, E)] = 0. \tag{20}$$

- For example, if we are considering a two-band model, the characteristic polynomial has the following form,

$$f(\beta, E) = E^2 + a(\beta)E + b(\beta), \qquad f(\beta e^{i\theta}, E) = E^2 + a(\beta e^{i\theta})E + b(\beta e^{i\theta}). \tag{21}$$

According to the result of Example A.8, their resultant relative to $E$ can be calculated directly.

Now we eliminate $\theta$. Before that, we need to clarify some properties of $G(\beta, \theta)$ in Eq. 20. Obviously, it is a complex algebraic function of $\beta$ and $\beta e^{i\theta}$, which can be represented by two independent real algebraic equations, namely,

$$G_r(\beta, \theta) := \operatorname{Re} G(\beta, \theta) = 0, \qquad G_i(\beta, \theta) := \operatorname{Im} G(\beta, \theta) = 0. \tag{22}$$

According to $\beta e^{i\theta} = \beta \cos\theta + i\beta \sin\theta$, $G_r(\beta, \theta)$ and $G_r(\beta, \theta)$ are two algebraic functions of $\cos\theta$ and $\sin\theta$, but not $\theta$. This means $\theta$ can not be directly eliminated based on the resultant method. In order to eliminate $\theta$, we need to use the Weierstrass substitution

$$\cos\theta = (1 - t^2)/(1 + t^2), \qquad \sin\theta = 2t/(1 + t^2). \tag{23}$$

Combining them with the resultant method, $t$ can be eliminated and the constraint equation of the aGBZ can be finally obtained,

$$F_{aGBZ}(\operatorname{Re}\beta, \operatorname{Im}\beta) := R_t[G_r(\beta, t), G_i(\beta, t)] = 0. \tag{24}$$

From the above results, $F_{aGBZ}(\operatorname{Re}\beta, \operatorname{Im}\beta)$ must be a real algebraic polynomial of $\operatorname{Re}\beta$ and $\operatorname{Im}\beta$. Here we note that if two polynomials

$$f(x) = f_0(x)f_1(x), \qquad g(x) = f_0(x)g_1(x) \tag{25}$$

have a common factor, their resultant must be trivial. In order to have a nontrivial result, we can calculate the resultant of $f_1(x)$ and $g_1(x)$. In the last content of the SM, we have provided a *Mathematica* code to show how to use the above procedure to calculate the aGBZ in a concrete example.

### C. Some examples

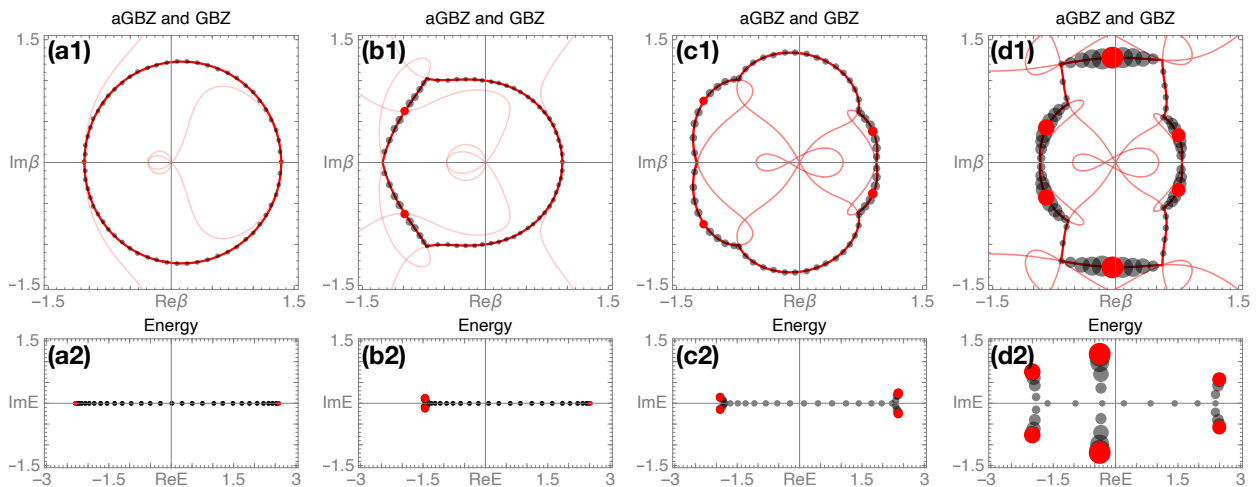

FIG. 3: The GBZ, aGBZ and the corresponding open boundary energy spectrum of the Hamiltonian determined by the characteristic equation Eq. (26). In (a1)-(d1), the red curves, gray points, and red points represent the aGBZ, numerical calculated GBZ with ($N = 30$), and self-conjugate points satisfying $\beta_p = \beta_{p+1}$. The size of the dots is proportional to $|\operatorname{Im} E|$. The analytic arcs containing the red points in the aGBZ must form the GBZ. The energy spectrum with $N = 30$ are plotted in the second row. The model is shown in Eq. (26) and the parameters are chosen to be $t_{-2} = 1/5, t_{-1} = 3/2, t_1 = 1$ for (a); $t_{-2} = 1/5, t_{-1} = 1, t_1 = 1, t_2 = 1/3$ for (b); $t_{-3} = -1/2, t_{-2} = 1/5, t_{-1} = 3/2, t_1 = 1, t_2 = 1/6$ for (c); $t_{-3} = -1/2, t_{-2} = 1/5, t_{-1} = 3/2, t_1 = 1, t_2 = 1/6, t_3 = 1/2$ for (d).

Here we will provide some additional examples for the single band model, where the characteristic equation is

$$f(\beta, E) = E - \sum_{m=-p}^{s} t_m \beta^m. \tag{26}$$

Fig. 3 shows some examples of aGBZ and GBZ. The nonzero parameters are chosen to be $t_{-2} = 1/5, t_{-1} = 3/2, t_1 = 1$ for (a); $t_{-2} = 1/5, t_{-1} = 1, t_1 = 1, t_2 = 1/3$ for (b); $t_{-3} = -1/2, t_{-2} = 1/5, t_{-1} = 3/2, t_1 = 1, t_2 = 1/6$ for (c); $t_{-3} = -1/2, t_{-2} = 1/5, t_{-1} = 3/2, t_1 = 1, t_2 = 1/6, t_3 = 1/2$ for (d). As a comparison, we also plot the numerical calculation of the GBZ in (a1)-(d1) with the gray points, whose size is proportional to $|\operatorname{Im} E|$. Any arcs containing the self-conjugate points satisfying $\beta_p = \beta_{p+1}$ (red points) must form the GBZ. The corresponding numerical eigenvalues are shown in (a2)-(d2)

### D. Discussions about auxiliary GBZ theory

The aGBZ theory can also be generalized to higher dimensional systems. For example, the aGBZ defined in 2D is the projection of the following equations $f(\beta_x, \beta_y, E) = f(\beta_x e^{i\theta_x}, \beta_y, E) = f(\beta_x, \beta_y e^{i\theta_y}, E) = 0$ on the $\beta_x$-$\beta_y$-plane, where $\beta_{x/y} = e^{ik_{x/y}}$ and $\theta_{x/y} \in \mathbb{R}$. For the interacting systems, our theory can also be extended by adding the self-energy correction to the noninteracting Hamiltonian, i.e., $\mathcal{H}_{eff}(k) = \mathcal{H}_0(k) + \Sigma(k, \omega = 0)$.

## III. SOME DETAILS OF THE SINGLE-BAND MODEL IN THE MAIN TEXT

In this section, we provide some details of the single-band model in the main text, whose non-Bloch Hamiltonian is

$$\mathcal{H}(\beta) = -1/6 - 1/(2\beta^3) + 8/(5\beta^2) + 10/(3\beta) + 4\beta + 2\beta^2 + \beta^3. \tag{27}$$

### A. Numerical result of Fig. 2

In this subsection, we show some additional numerical results of Fig. 2 in the main text. In Fig. 4(a), we numerically solve the open boundary Hamiltonian with size $N = 2000$ and precision $P = 800$ by *Mathematica* software. The calculation time is about 3 days. In Fig. 4(b), we numerically solve the same Hamiltonian with $N = 3000$ and precision $P = 1800$ by *Mathematica*. The calculation time is about 11 days. As discussed in the main text, even under the parameters calculated in Fig. 4(b), we still do not know whether there exist self-intersection points on the GBZ. Additionally, the result with $N = 200$ and $P = 8$ (machine precision) is shown in Fig. 4(c), from which we can see that the numerical calculation results become unreliable under the current calculation precision.

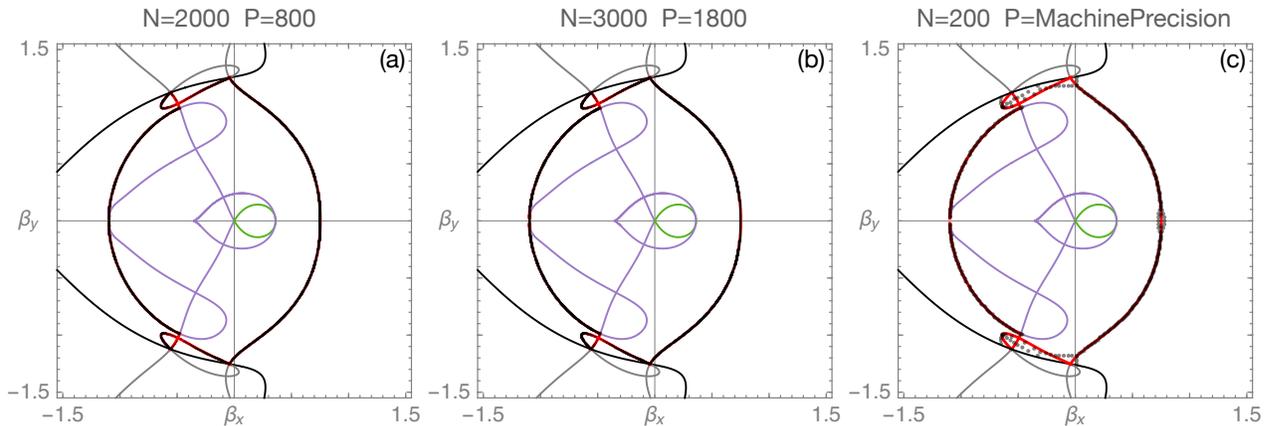

FIG. 4: The numerical (dots) and analytical (lines) results of Fig. 1 in the main text. Here $\beta_x := \operatorname{Re}\beta$, $\beta_y := \operatorname{Im}\beta$. The calculation time for obtaining (b) is 11 days. The higher precision is necessary, otherwise, the numerical calculation will have a large error as illustrated in (c), in which the $P$ is taken as machine precision ($P = 8$).

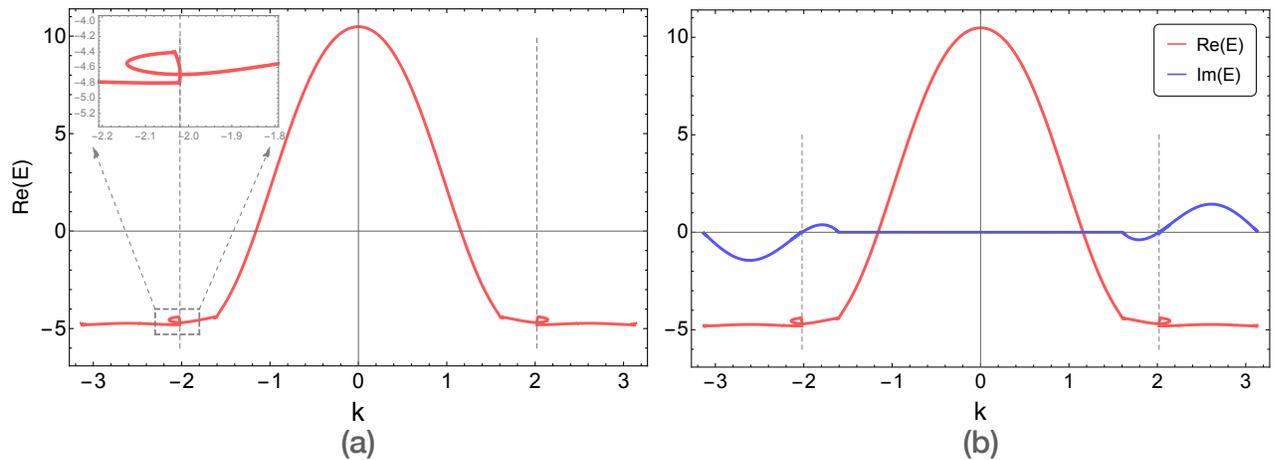

FIG. 5: (a) shows the real part of $E(\beta_{GBZ}(k))$ of the model Eq. 27. The vertical dashed lines represent the self-intersections, near which the enlarged view of band structure is plotted in the subfigure. (b) shows the comparison of the real (red line) and imaginary (blue line) parts of $E(\beta_{GBZ}(k))$.

## B. The physics of self-intersections on GBZ

As mentioned in the main text, the self-intersection points in the single-band model can be regarded as a new kind of singularity point, which has no counterpart in Hermitian systems and is unique to non-Hermitian systems. Now we explain this. Review that in one-dimensional Hermitians, the Van Hove singularity $k_s$ is determined by

$$\text{Hermitain}: \quad \partial_k E_\mu(k_s) = 0, \quad \mu = 1, ..., n \tag{28}$$

Generalizing the above concept to the non-Hermitian Hamiltonians, from $k = -i \ln \beta$, we know that $k$ can be regarded as the argument of $\beta_{GBZ}$, which is labeled by $\beta_{GBZ}(k)$. Consider a general multi-band non-Hermitian Hamiltonian $\mathcal{H}(\beta)$, the corresponding eigenvalues are $\{E_\mu(\beta), \mu = 1, ..., n\}$, where $n$ is the number of energy bands. When open boundary condition is chosen, the corresponding continuous band spectrum are determined by

$$\{E_\mu, \mu = 1, ..., n\} = \{E_\mu(\beta_{GBZ,\mu})\} \qquad \to \qquad E_\mu(k) = E_\mu(\beta_{GBZ,\mu}(k)), \tag{29}$$

where $\beta_{GBZ,\mu}$ is used to represent the GBZ of the $\mu$-th band. Therefore, the singularity point Eq. 28 can be generalized to non-Hermitian bands as follows

$$\text{non} - \text{Hermitain}: \quad \partial_k E_\mu(\beta_{GBZ,\mu}(k_s)) = 0. \tag{30}$$

Back to our example Eq. 27, whose GBZ band structures are shown in Fig. 5. At the (single-band) self-intersection point on the GBZ, since $\partial \beta_{GBZ}(k_s)/\partial k = 0$, $k_s$ must also satisfy the condition Eq. 30, which means the self-intersection is a new type of non-Hermitian singularity point, which has no counterpart in Hermitian systems and is unique to non-Hermitian systems. In addition, we note that the self-conjugate point on the GBZ satisfying $\beta_p = \beta_{p+1}$ corresponds to the conventional Van Hove singularity in the Hermitian case.

## IV. WAVEFUNCTION WINDING NUMBER IN THE MULTIPLE SUB-GBZ CASES

In the main text, we have mentioned that due to the existence of multiple sub-GBZs, the definitions of the wave-function winding number in Ref. [1] and Ref. [2] can not be extended directly. We further pointed out that our conjectured formula Eq. 9 (in the main text) can reduce to the formulas in the multi-band Hermitian and two-band non-Hermitian cases. In this section, we will explain and prove the above statements.

## A. Wavefunction winding number in the presence of sub-lattice symmetry

Consider a general $2m$-band Bloch Hamiltonian preserving sub-lattice symmetry, $\mathcal{S}\mathcal{H}(k)\mathcal{S}^{-1} = -\mathcal{H}(k)$. When the sub-lattice symmetry is represented by $\mathcal{S} = \tau_z$, the corresponding Hamiltonian can be written as the following form,

$$\mathcal{H}(k) = \begin{pmatrix} 0 & R_+(k) \\ R_-(k) & 0 \end{pmatrix}. \tag{31}$$

When $R_-(k) = R_+^\dagger(k)$, the Hamiltonian becomes Hermitian, otherwise, it is non-Hermitian. We note that the eigenvalues of Eq. 31 can be obtained from the following eigenequation

$$R_+(k)R_-(k)\left|a_R(k)\right\rangle = E_\mu^2(k)\left|a_R(k)\right\rangle. \tag{32}$$

To be more precise, if $E_\mu^2(k)$ is an eigenvalue of $R_+(k)R_-(k)$, then, $\pm E_\mu(k)$ are the eigenvalues of $\mathcal{H}(k)$. The wavefunction winding number can be defined through the $Q$ matrix [1], whose formula has been derived from the SM of Ref. [2],

$$Q(k) = \sum_{\mu=1}^{m} \frac{1}{E_\mu(k)} \begin{pmatrix} O & \left|a_\mu^R(k)\right\rangle \left\langle a_\mu^L(k)\right| R_+(k) \\ R_-(k)\left|a_\mu^R(k)\right\rangle \left\langle a_\mu^L(k)\right| & O \end{pmatrix}, \tag{33}$$

where

$$\left\langle a_L(k)\right| R_+(k)R_-(k) = E_\mu^2(k)\left\langle a_L(k)\right| \;, \quad \left\langle a_\alpha^L(k) | a_\beta^R(k)\right\rangle = \delta_{\alpha\beta}. \tag{34}$$

We note that the above equations can be generalized to the non-Bloch Hamiltonians $\mathcal{H}(\beta)$ by replacing $e^{ik}$ by $\beta$.

### 1. Two-band case

For the two-band model, the $Q$ matrix in the non-Bloch form becomes

$$Q(\beta) = \frac{1}{E(\beta)} \begin{pmatrix} 0 & R_+(\beta) \\ R_-(\beta) & 0 \end{pmatrix}. \tag{35}$$

Define

$$q(\beta) = \frac{R_+(\beta)}{\sqrt{R_+(\beta)R_-(\beta)}} = \sqrt{\frac{R_+(\beta)}{R_-(\beta)}}. \tag{36}$$

Now, the wavefunction winding number becomes

$$w = \frac{1}{2\pi i} \oint_{\beta_{GBZ}} dq\, q^{-1}(\beta) = \frac{i}{2\pi} \oint_{\beta_{GBZ}} d\ln q(\beta) = \frac{1}{2}(w_+ - w_-), \tag{37}$$

where

$$w_\pm = \frac{1}{2\pi i} \oint_{\beta_{GBZ}} d\ln R_\pm(\beta) = -P_\pm + Z_\pm. \tag{38}$$

Here $Z_\pm$ are the number of zeros note only satisfying $R_\pm(\beta) = 0$ but also being inside the GBZ $\beta_{GBZ}$, and $P_\pm$ are the orders of the pole of $R_\pm(\beta)$. One can notice that this formula is equivalent to Eq. 9 in the main text.

### 2. Four-band case

If $m = 2$, both $R_+(\beta)$ and $R_-(\beta)$ are $2 \times 2$ matrix. The corresponding eigenvalues can be labeled by $\pm E_1(\beta)$ and $\pm E_2(\beta)$. Therefore, the $Q$ matrix becomes

$$Q(\beta) = \sum_{\mu=1}^{2} \frac{1}{E_\mu(\beta)} \begin{pmatrix} O & \left|a_\mu^R(\beta)\right\rangle \left\langle a_\mu^L(\beta)\right| R_+(\beta) \\ R_-(\beta)\left|a_\mu^R(\beta)\right\rangle \left\langle a_\mu^L(\beta)\right| & O \end{pmatrix}. \tag{39}$$

Define

$$q(\beta) = \sum_{\mu=1}^{2} q_\mu(\beta) := \sum_{\mu=1}^{2} \frac{1}{E_\mu(\beta)} \left| a_\mu^R(\beta) \right\rangle \left\langle a_\mu^L(\beta) \right| R_+(\beta). \tag{40}$$

We have two different cases in the following discussion.

- (i) $\beta_{GBZ,1} = \beta_{GBZ,2}$:

  When the sub-GBZs for $E_1(\beta)$ and $E_2(\beta)$ are the same, e.g., in the Hermitian case, the wining number is defined as

  $$w = \frac{1}{2\pi i} \oint_{\beta_{GBZ}} \text{Tr} \left[ dq q^{-1}(\beta) \right] = \frac{1}{2\pi i} \oint_{\beta_{GBZ}} d\,\text{Tr}\left[ \ln q(\beta) \right] = \frac{1}{2\pi i} \oint_{\beta_{GBZ}} d \ln \det \left[ q(\beta) \right]. \tag{41}$$

- (ii) $\beta_{GBZ,1} \neq \beta_{GBZ,2}$:

  However, when the sub-GBZs for $E_1(\beta)$ and $E_2(\beta)$ are different, it is not clear which sub-GBZs should be chosen as the integration path, because the $q(\beta)$ is defined for all the bands with negative eigenvalues, e.g., $-E_1(\beta)$ and $-E_2(\beta)$. Naively, we may expect the following formula works,

  $$w \overset{?}{=} \frac{1}{2\pi i} \sum_{\mu=1}^{2} \oint_{\beta_{GBZ,\mu}} \text{Tr} \left[ dq_\mu q_\mu^{-1}(\beta) \right] = \frac{1}{2\pi i} \sum_{\mu=1}^{2} \oint_{\beta_{GBZ,\mu}} d \ln \det \left[ q_\mu(\beta) \right], \tag{42}$$

  where

  $$q_\mu(\beta) = \frac{1}{E_\mu(\beta)} \left| a_\mu^R(\beta) \right\rangle \left\langle a_\mu^L(\beta) \right| R_+(\beta) \tag{43}$$

  and $\beta_{GBZ,\mu}$ are the sub-GBZs for the $-E_\mu(\beta)$ band. However, this formula is not correct. The reason is that it can not reduce to the correct one in the $\beta_{GBZ,1} = \beta_{GBZ,2}$ case due to

  $$\det \left[ \sum_{\mu=1}^{2} q_\mu(\beta) \right] \neq \sum_{\mu=1}^{2} \det \left[ q_\mu(\beta) \right]. \tag{44}$$

This subsection (i) explains why the definitions of the wavefunction winding number in Ref. [1] and Ref [2] can not be extended directly in the multiple sub-GBZ cases; (ii) proves our conjectured formula Eq. 9 (in the main text) can reduce to the formula in two-band non-Hermitian cases.

## B. Reduce to the Hermitian case

In this subsection, we will show that our conjectured formula Eq. 9 (in the main text) can reduce to the formula in the multi-band Hermitian case.

In the Hermitian case, we have

$$w_\pm = \frac{1}{2\pi i} \oint_{\beta_{BZ}} d \ln \det[R_\pm(\beta)] = -P_\pm + Z_\pm, \tag{45}$$

where $Z_\pm$ are the number of zeros note only satisfying $R_\pm(\beta) = 0$ but also being inside the BZ $\beta_{BZ} = e^{ik}$, and $P_\pm$ are the orders of the pole of $R_\pm(\beta)$. Since in the Hermitian case, $R_+(k) = R_-^\dagger(k)$, we have $\det[R_+(\beta)] = \det[R_-(\beta)]^*$. As a result, we have

$$w_+ + w_- = 0. \tag{46}$$

In the Sup Mat of Ref. [8], the authors proved that the wavefunction winding number in the Hermitian case can be written as

$$w = \frac{1}{2\pi i} \oint_{\beta_{BZ}} d \ln \det[R_+(\beta)] = w_+. \tag{47}$$

Combining Eq. 46 and Eq. 47, we finally have

$$w = \frac{1}{2}(w_+ - w_-). \tag{48}$$

This formula is equivalent to Eq. 9 in the main text.

## V. MATHEMATICA CODE

# Methods

## The single – band Hamiltonian and characteristic equation

```
In[1]:= Clear["Global`*"]

In[2]:= {t1, t2, t3, w1, w2, w3} = {3 / 2, 1 / 5, -5 / 10, 1, 1 / 6, 5 / 10};
      hamz[z_] := t1 / z + t2 / z^2 + t3 / z^3 + w1 z + w2 z^2 + w3 z^3;
      hamr[dim_] := Normal[
        SparseArray[{Band[{2, 1}, {dim, dim}] → {t1}, Band[{3, 1}, {dim, dim}] → {t2},
          Band[{4, 1}, {dim, dim}] → {t3}, Band[{1, 2}, {dim, dim}] → {w1},
          Band[{1, 3}, {dim, dim}] → {w2}, Band[{1, 4}, {dim, dim}] → {w3}}, {dim, dim}]]
      eigval[dim_] := Eigensystem[N[hamr[dim], 32]][[1]]
      f[e_, z_] := e - hamz[z]
```

## Eigenvalues of open system with finite size

```
In[7]:= P0 = ListPlot[ReIm[eigval[100]], PlotStyle → {PointSize[0.015], Brown},
       Frame → True, FrameLabel → {"Re(E)", "Im(E)"}, PlotRange → All]
```

Out[7]=

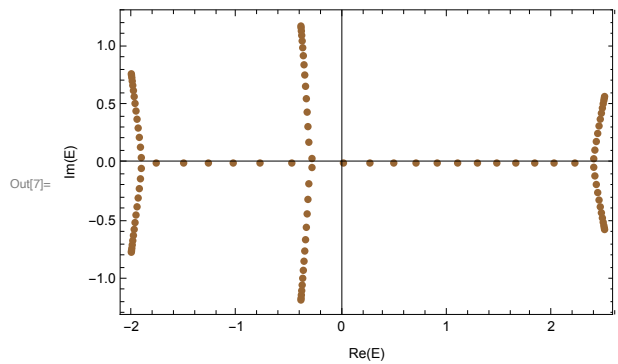

## Auxiliary generalized Brillouin zone

```
In[8]:=  RealRes = Factor[ComplexExpand[
            Re[Resultant[f[Ene, x + I y], f[Ene, (x + I y) ((1 - t^2 + 2 I t) / (1 + t^2))]], Ene]]]]
        ImagRes = Factor[ComplexExpand[
            Im[Resultant[f[Ene, x + I y], f[Ene, (x + I y) ((1 - t^2 + 2 I t) / (1 + t^2))]], Ene]]]]
```

$$Out[8]= \frac{1}{15\left(1+t^2\right)^3\left(x^2+y^2\right)^3}$$

t (t x + y) (−135 x² + 90 t² x² − 15 t⁴ x² + 24 x³ + 24 t² x³ + 45 x⁴ + 90 t² x⁴ + 45 t⁴ x⁴ +
    30 x⁶ + 60 t² x⁶ + 30 t⁴ x⁶ + 20 x⁷ + 20 t² x⁷ + 135 x⁸ − 90 t² x⁸ + 15 t⁴ x⁸ + 360 t x y −
    120 t³ x y − 24 t x² y − 24 t³ x² y − 20 t x⁶ y − 20 t³ x⁶ y − 360 t x⁷ y + 120 t³ x⁷ y +
    45 y² − 150 t² y² + 45 t⁴ y² + 24 x y² + 24 t² x y² + 90 x² y² + 180 t² x² y² + 90 t⁴ x² y² +
    90 x⁴ y² + 180 t² x⁴ y² + 90 t⁴ x⁴ y² + 60 x⁵ y² + 60 t² x⁵ y² + 360 x⁶ y² − 120 t² x⁶ y² −
    24 t y³ − 24 t³ y³ − 60 t x⁴ y³ − 60 t³ x⁴ y³ − 1080 t x⁵ y³ + 360 t³ x⁵ y³ + 45 y⁴ +
    90 t² y⁴ + 45 t⁴ y⁴ + 90 x² y⁴ + 180 t² x² y⁴ + 90 t⁴ x² y⁴ + 60 t² x³ y⁴ +
    270 x⁴ y⁴ + 180 t² x⁴ y⁴ − 90 t⁴ x⁴ y⁴ − 60 t x² y⁵ − 60 t³ x² y⁵ − 1080 t x³ y⁵ +
    360 t³ x³ y⁵ + 30 y⁶ + 60 t² y⁶ + 30 t⁴ y⁶ + 20 x y⁶ + 20 t² x y⁶ + 360 t² x² y⁶ −
    120 t⁴ x² y⁶ − 20 t y⁷ − 20 t³ y⁷ − 360 t x y⁷ + 120 t³ x y⁷ − 45 y⁸ + 150 t² y⁸ − 45 t⁴ y⁸)

$$Out[9]= \frac{1}{15\left(1+t^2\right)^3\left(x^2+y^2\right)^3}$$

t (−45 x³ + 150 t² x³ − 45 t⁴ x³ + 12 x⁴ − 12 t⁴ x⁴ + 45 x⁵ + 90 t² x⁵ + 45 t⁴ x⁵ − 30 x⁷ − 60 t² x⁷ −
    30 t⁴ x⁷ − 10 x⁸ + 10 t⁴ x⁸ − 45 x⁹ + 150 t² x⁹ − 45 t⁴ x⁹ + 405 t x² y − 270 t³ x² y +
    45 t⁵ x² y − 48 t x³ y − 48 t³ x³ y − 45 t x⁴ y − 90 t³ x⁴ y − 45 t⁵ x⁴ y + 30 t x⁶ y +
    60 t³ x⁶ y + 30 t⁵ x⁶ y + 40 t x⁷ y + 40 t³ x⁷ y + 405 t x⁸ y − 270 t³ x⁸ y + 45 t⁵ x⁸ y +
    135 x y² − 450 t² x y² + 135 t⁴ x y² + 90 x³ y² + 180 t² x³ y² + 90 t⁴ x³ y² − 90 x⁵ y² −
    180 t² x⁵ y² − 90 t⁴ x⁵ y² − 20 x⁶ y² + 20 t⁴ x⁶ y² − 135 t y³ + 90 t³ y³ − 15 t⁵ y³ −
    48 t x y³ − 48 t³ x y³ − 90 t x² y³ − 180 t² x² y³ − 90 t⁵ x² y³ + 90 t x⁴ y³ + 180 t³ x⁴ y³ +
    90 t⁵ x⁴ y³ + 120 t x⁵ y³ + 120 t³ x⁵ y³ + 1080 t x⁶ y³ − 720 t³ x⁶ y³ + 120 t⁵ x⁶ y³ −
    12 y⁴ + 12 t⁴ y⁴ + 45 x y⁴ + 90 t² x y⁴ + 45 t⁴ x y⁴ − 90 x³ y⁴ − 180 t² x³ y⁴ − 90 t⁴ x³ y⁴ +
    270 x⁵ y⁴ − 900 t² x⁵ y⁴ + 270 t⁴ x⁵ y⁴ − 45 t y⁵ − 90 t³ y⁵ − 45 t⁵ y⁵ + 90 t x² y⁵ +
    180 t³ x² y⁵ + 90 t⁵ x² y⁵ + 120 t x³ y⁵ + 120 t³ x³ y⁵ + 810 t x⁴ y⁵ − 540 t³ x⁴ y⁵ +
    90 t⁵ x⁴ y⁵ − 30 x y⁶ − 60 t² x y⁶ − 30 t⁴ x y⁶ + 20 x² y⁶ − 20 t⁴ x² y⁶ + 360 x³ y⁶ −
    1200 t² x³ y⁶ + 360 t⁴ x³ y⁶ + 30 t y⁷ + 60 t³ y⁷ + 30 t⁵ y⁷ + 40 t x y⁷ + 40 t³ x y⁷ +
    10 y⁸ − 10 t⁴ y⁸ + 135 x y⁸ − 450 t² x y⁸ + 135 t⁴ x y⁸ − 135 t y⁹ + 90 t³ y⁹ − 15 t⁵ y⁹)

## Eliminate the common factors

```
In[10]:=  aGBZ = Resultant[RealRes * (t² + 1)³ (x² + y²)³ / t, ImagRes * (t² + 1)³ (x² + y²)³ / t, t];
          {FactorList[aGBZ][[;; , 1]], FactorList[aGBZ][[;; , 2]]};
          {aGBZ1, aGBZ2} = {FactorList[aGBZ][[4, 1]], FactorList[aGBZ][[5, 1]]};
```

## Plot aGBZ

FIG. 7: Mathematica code

In[13]:= `P1 = ContourPlot[{aGBZ1 ⩵ 0, aGBZ2 ⩵ 0}, {x, -1.5, 1.5}, {y, -1.5, 1.5},`
   `PlotPoints → 100, ContourStyle → {{Darker[Red], Thickness[0.008]},`
      `{Darker[Blue], Thickness[0.008]}}, PlotTheme → "Scientific",`
   `FrameLabel → {{HoldForm["Re(β)"], None}, {HoldForm["Im(β)"], None}}]`

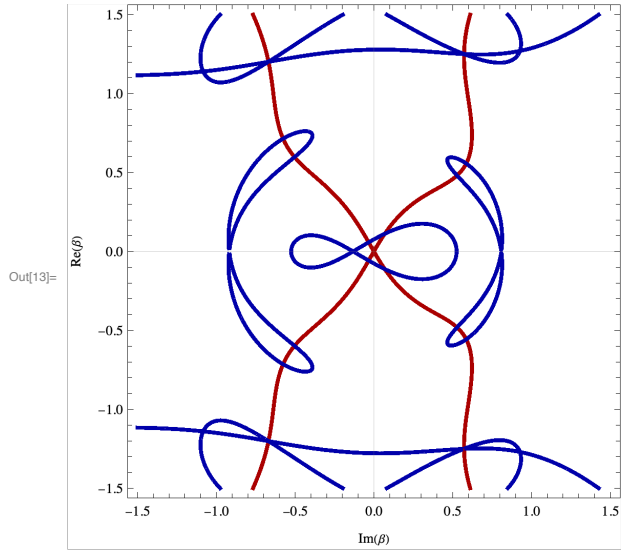

**Pick critical points**

FIG. 8: Mathematica code

```
In[14]:= dege = e /. NSolve[{f[e, z], D[f[e, z], z]} == 0, {e, z}];
       degz1 = Table[Table[Sort[z /. NSolve[f[dege[[i]], z], z], Abs[#1] < Abs[#2] &],
            {i, 1, Length[dege]}][[j]][[3]], {j, 1, Length[dege]}];
       degz2 = Table[Table[Sort[z /. NSolve[f[dege[[i]], z], z], Abs[#1] < Abs[#2] &],
            {i, 1, Length[dege]}][[j]][[4]], {j, 1, Length[dege]}];
       degz1 - degz2
       CPoint = Graphics[Table[{Red, PointSize[0.015 + Abs[hamz[degz1[[i]]]] / 100],
            Point[{Re[degz1[[i]]], Im[degz1[[i]]]}]}, {i, 1, 6}]];
       Show[
         {P1,
          CPoint}]
```

Out[17]= {0. + 0. i, 0. + 0. i, 0. + 0. i, 0. + 0. i, 0. + 0. i, 0. + 0. i}

Out[19]=

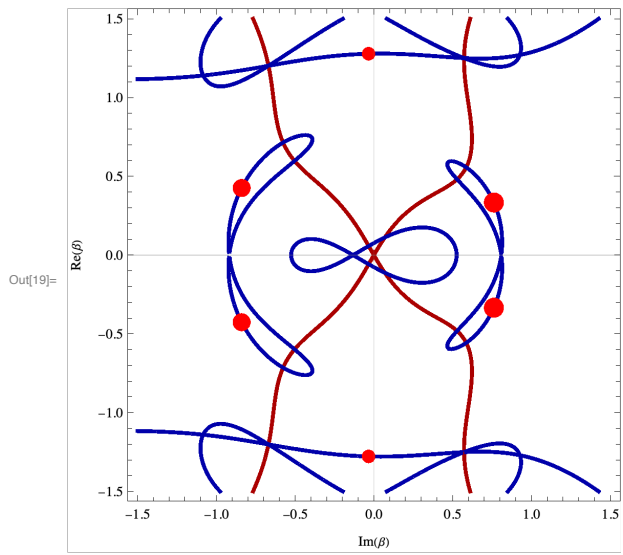

### Compare with exact solutions

```
In[20]:= BZ1 = Table[Sort[z /. NSolve[f[eigval[30][[i]], z], z], Abs[#1] < Abs[#2] &][[3]],
            {i, 1, 30}];
       BZ2 = Table[Sort[z /. NSolve[f[eigval[30][[i]], z], z], Abs[#1] < Abs[#2] &][[4]],
            {i, 1, 30}];
```

```
In[22]:= GBZpoint = Graphics[
         {Table[{Black, Opacity[0.4], PointSize[0.015 + Abs[hamz[BZ1[[i]]]] / 100],
            Point[{Re[BZ1[[i]]], Im[BZ1[[i]]]}]}, {i, 1, 30}],
          Table[{Black, Opacity[0.4], PointSize[0.015 + Abs[hamz[BZ2[[i]]]] / 100],
            Point[{Re[BZ2[[i]]], Im[BZ2[[i]]]}]}, {i, 1, 30}]}];
```

FIG. 9: Mathematica code

In[23]:= **Show[{P1, GBZpoint, CPoint}]**

Out[23]= 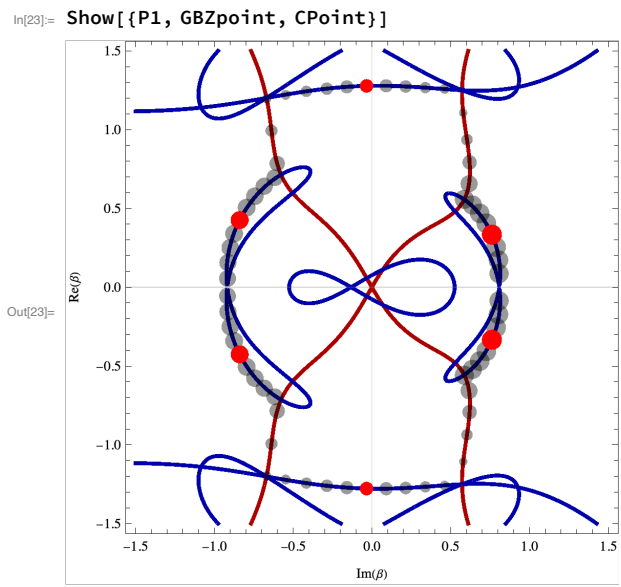



\end{document}